\PassOptionsToPackage{dvipsnames}{xcolor}
\documentclass[conference]{IEEEtran}
\IEEEoverridecommandlockouts
\usepackage{booktabs}
\usepackage[ruled,vlined,linesnumbered]{algorithm2e}
\usepackage{amssymb,amsmath,epsfig,latexsym,graphicx,bm}
\usepackage{ifpdf}
\ifpdf
  \usepackage{epstopdf}
\fi
\usepackage{xcolor}
\usepackage{soul}
\usepackage{cite}
\usepackage{amsfonts}
\usepackage{amsmath}
\usepackage{amsmath,amssymb}
\usepackage{textcomp}
\usepackage{xcolor}
\usepackage{optidef}
\usepackage[font=footnotesize]{caption}
\usepackage{acronym}

\acrodef{AP}{access point}
\acrodef{AoA}{angle of arrival}
\acrodef{ToA}{time of arrival}
\acrodef{ISAC}{Integrated sensing and communication}
\acrodef{SN}{subnetwork}
\acrodef{RCS}{radar cross section}
\acrodef{IIoT}{industrial internet of things}
\acrodef{MTI}{moving target indication}
\acrodef{CSI}{channel state information}
\acrodef{NMS}{non-maximum suppression}
\acrodef{CFAR}{constand false alarm rate}
\acrodef{OFDMA}{orthogonal frequency-division multiple access}
\acrodef{ES}{edge server}
\acrodef{eCRLB}{empirical Cramér–Rao lower bound}
\acrodef{PEB}{position error bound}
\acrodef{ULA}{uniform linear array}
\acrodef{WLS}{weighted least square}
\acrodef{FoV}{field of view}

\definecolor{amaranth}{rgb}{0.9, 0.17, 0.31}

\def\BibTeX{{\rm B\kern-.05em{\sc i\kern-.025em b}\kern-.08em
		T\kern-.1667em\lower.7ex\hbox{E}\kern-.125emX}}
\usepackage[font=footnotesize,caption=false]{subfig}

\title{Cooperative Multistatic Target Localization for ISAC in Cluttered Industrial IoT Networks}
\author{\IEEEauthorblockN{Mostafa Nozari, Israel Leyva-Mayorga, and Gilberto Berardinelli}  
\IEEEauthorblockA{Department of Electronic Systems, Aalborg University, Denmark (\{mnozari, ilm, gb\}@es.aau.dk)} 
}% \thanks{Corresponding author M. Nozari. Emails: \IEEEauthorrefmark{1}\{mnozari, ilm, gb\}@es.aau.dk, \IEEEauthorrefmark{2}fabio.saggese@ing.unipi.it. This work is supported by Independent Research Fund Denmark, grant no. 3105-00077B. The work of F. Saggese is supported by Horizon Europe MSCA Postdoctoral Fellowships with Grant~101204088.}%

\begin{document}
\maketitle
\bstctlcite{IEEEexample:BSTcontrol}
\begin{abstract}
In this paper, we propose a novel integrated sensing and communication (ISAC) framework for collaborative multistatic target localization in dense Industrial Internet-of-Things (IIoT) environments in the presence of environmental clutter. We first develop a lightweight temporal clutter-suppression learning method to mitigate persistent reflections. Building on this, we propose an iterative localization algorithm that integrates two key components introduced in this work: a sampling-based field-of-view-aware initialization (SFI) scheme and an empirical position error bound (PEB) scheme,
%based node selection strategy, 
which together adaptively identify the most informative subset of sensing nodes. A reliability-aware weighted least-squares estimator is then employed to 
%jointly 
fuse range and angle-of-arrival measurements from the selected sensing receivers for target localization. %Numerical results demonstrate rapid convergence of the proposed method, with the localization error decreasing by more than two orders of magnitude within six sensing iterations, while significantly outperforming all %realizable 
%considered benchmarks under the same sensing-resource budget. %The study further quantifies the impact of the number of antennas and multistatic receivers and highlights the trade-off among localization accuracy, communication cost, and localization latency
Numerical results demonstrate rapid convergence of the proposed method, reducing the localization RMSE by nearly two orders of magnitude within six sensing iterations to about \(\mathbf{45}\,\mathbf{cm}\), while significantly outperforming all considered benchmarks under the same sensing-resource budget.
\end{abstract}
\section{Introduction}
In the 6G era, in-X \acp{SN} are autonomous wireless networks embedded within physical entities, such as vehicles, robots, and production modules, where they support communication among internal components. In mission-critical %environments, 
%such as 
industrial %factories, 
environments,
their dense deployment makes it challenging to satisfy stringent latency and reliability requirements under a high degree of autonomy. Therefore, they must combine robust standalone operation with seamless integration into the  6G infrastructure for coordination, computation offloading, and resource management~\cite{berardinelli2021extreme}.

% Integrated sensing and communication (ISAC) has been
% considered a promising technology for the next generation wireless networks as it exploits the dual use of communication spectrum, hardware, and signaling for both data transmission and environmental sensing. In dense 5G/6G deployments, such a unified infrastructure can evolve into a perceptive network, where spatially distributed nodes, such as in factory IIoT subnetworks, collaboratively provide diverse observation perspectives for environmental awareness tasks, including detection, localization, tracking, and mapping. %This capability is particularly promising for dense in-X subnetwork deployments, as it supports not only reliable internal connectivity but also efficient cooperative sensing of the surrounding environment.
\ac{ISAC} has emerged as a promising technology for next-generation wireless networks by enabling  dual use of spectrum, hardware, and signaling for both data transmission and environmental sensing. In dense 5G/6G deployments, such a unified infrastructure can evolve into a perceptive network, where spatially distributed nodes, such as \ac{IIoT} \acp{SN}, collaboratively provide diverse observation perspectives for tasks, including detection, localization, tracking, and mapping.

Cooperative sensing, especially in scenarios with spatially distributed nodes, has recently emerged as a promising research direction in \ac{ISAC}, where measurements from multiple nodes are utilized to improve sensing accuracy~\cite{nozari2025toward,Li2025cooperative}. However, adaptively selecting a limited subset of nodes that provides rich geometric diversity from a large pool of candidates remains an open challenge.

Recent SN studies have mainly focused on communication aspects such as resource management and interference coordination~\cite{du2022multi,abode2025goal}. %Only recently has the integration of subnetworks for radio sensing been considered in~\cite{nozari2025toward}, where the most informative subnetworks are selected to cooperatively sense the environment in a mono-static setting and subsequently share range measurements at the ES for target localization under both Gaussian and fading channels.
In our previous work~\cite{nozari2025toward}, %the integration of \acp{SN} for radio sensing was considered,
\acp{SN} were considered for radio sensing,
where informative \acp{SN} were selected to cooperatively sense the environment and share range measurements at the \ac{ES} for target localization under Gaussian and fading channels. However, this framework did not incorporate \ac{AoA} measurements and was not designed for \ac{RCS}-agnostic localization. %Moreover, the transmitter--receiver co-location constraint inherent to the mono-static setting limits geometric diversity and increases susceptibility to self-interference.
Moreover, the monostatic transmitter (Tx)--receiver (Rx) co-location constraint 
%inherent to
%of
%the mono-static setting 
limits geometric diversity and typically requires sophisticated self-interference cancelation.
%techniques.

Sensing in cluttered environments, with potentially strong interference scattered from surrounding objects, is challenging. While some of \ac{ISAC} studies  assumes idealized clutter-free sensing~\cite{behdad2022power}, other rely on clutter mitigation through separability assumptions and statistical signatures. A practical alternative is Doppler-domain \ac{MTI}/high-pass filtering~\cite{wang2024dynamic}. However, it can create blind Doppler regions.
 
%To bridge these gaps,
% This paper proposes a comprehensive multistatic sensing framework to extract the \ac{ToA} and \ac{AoA} from the most informative subset of subnetworks in cluttered industrial environments under imperfect \ac{CSI} for \ac{RCS}-agnostic target localization. 
This paper proposes a novel multistatic sensing framework for target localization in a dense cluttered \ac{IIoT} scenario under imperfect \ac{CSI}, where the most informative subset of collaborative \acp{SN} is dynamically selected to improve localization accuracy with limited latency. 
The main contributions are as follows:
\begin{itemize}
    \item We propose a novel cooperative sensing framework for ISAC operating in a bi/multistatic sensing configuration that is applicable to practical scenarios with environmental clutter and imperfect \ac{CSI}. 
    % \item We develop a simple yet effective clutter suppression method to mitigate unwanted background reflections at the preprocessing stage  based on temporal background estimation in the pre-processing stage. The proposed approach does not require geometric assumptions, clutter power models, or target motion assumptions.

    % \item We propose a low-complexity iterative algorithm to select an optimal subset of sensing subnetworks under limited delay and resource budgets. To further enhance localization robustness, we employ a weighted least squares (WLS) fusion scheme that accounts for the reliability of range–AoA measurements across the selected subnetworks.
    \item We develop a lightweight and model-agnostic clutter-suppression learning method based on temporal background estimation, which does not rely on explicit clutter-power modeling or motion-based filtering assumptions.

    % \item We design a coarse-to-fine sensing pipeline that combines range--angle power map construction, edge-aware 2D-CFAR detection to obtain coarse ToA and AoA estimates, and sub-resolution refinement via phase-slope delay estimation and MUSIC-based angular estimation. Moreover, a two-dimensional non-maximum suppression (NMS) step is incorporated to mitigate sidelobe effects and suppress multiple detections originating from the same peak region.
    
    % \item We propose a node-selection and reliability-aware fusion framework that combines sampling-based field-of-view (FoV)-aware initialization with an empirical \ac{PEB}-based node-selection criterion to maximize geometric visibility and adaptively identify the most informative sensing links. The \ac{ToA} and \ac{AoA} measurements from the selected links are then jointly fused via a \ac{WLS} estimator, where \ac{eCRLB} models are introduced to adaptively weight the reliability of each modality across sensing \acp{SN}.
    \item We propose a node-selection and reliability-aware fusion scheme that combines a sampling-based \ac{FoV}-aware initialization with empirical \ac{PEB}-based node selection. This approach maximizes geometric visibility and identifies the most informative sensing links. The selected links provide \ac{ToA} and \ac{AoA} measurements, which are jointly fused via a \ac{WLS} estimator. The \ac{eCRLB} models are used to  weight the reliability of each measurement.
    \item We evaluate the proposed method under clutter-free, clutter-suppressed, and cluttered conditions against theoretical and empirical benchmarks, and analyze localization accuracy and latency--communication trade-offs for varying antenna  and multistatic receiver counts.

\end{itemize}

\section{System Model}\label{sec:sysmodel}
\begin{figure}[t]
\centering
\includegraphics[width=1.00\columnwidth]{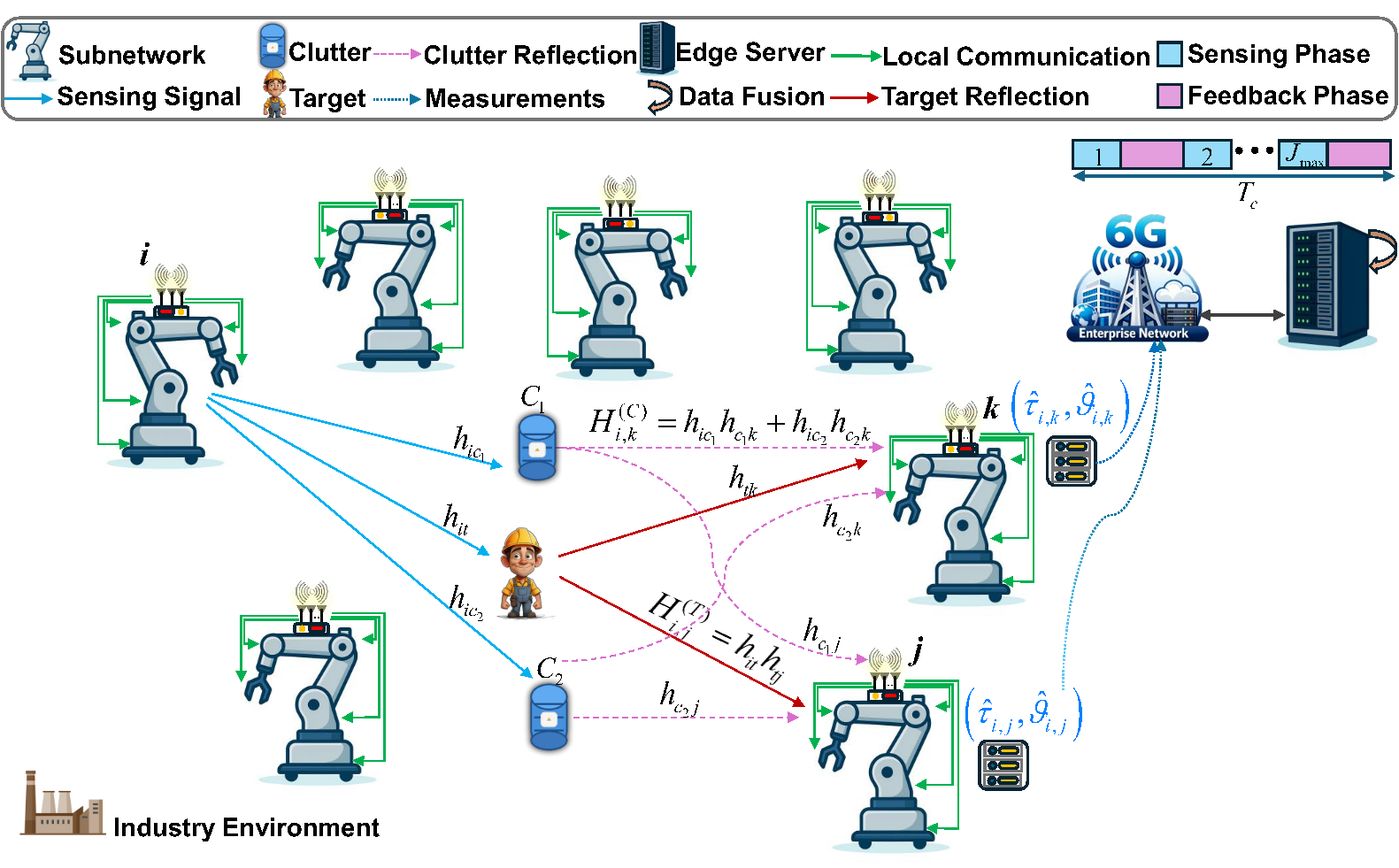}
\caption{Dense IIoT factory with $N$ \acp{SN} supporting local downlink communication while collaboratively localizing a target via multistatic OFDM sensing.}
\label{SYYYSSIIMO}
\vspace{-5mm}
\end{figure}

%We consider a dense Industrial Internet-of-Things (IIoT) network that supports joint downlink communication, such as closed-loop control and radar sensing over a shared spectrum%using OFDM modulation, as 
%, as illustrated in Fig.~\ref{SYYYSSIIMO}. %The OFDM waveform employs $N_{\mathrm{sc}}$ subcarriers with a frequency spacing of $\Delta f$, starting from the lowest frequency $f_0$. Accordingly, the total transmission bandwidth is given by $W = (N_{\mathrm{sc}} - 1)\Delta f$, and the frequency of the $k$-th subcarrier is expressed as $f_k = f_0 + k\Delta f$, where $k = 0, 1, \dots, N_{\mathrm{sc}} - 1$.}
We consider an \ac{IIoT} network consisting of an \ac{ES} and $N$ spatially distributed \acp{SN} with communication and sensing capabilities, which cooperate to enable on-demand target localization for monitoring and safety purposes, as illustrated in Fig.~\ref{SYYYSSIIMO}. 
The \ac{ES} orchestrates sensing-communication operations, manages resource allocation across \acp{SN}, and performs centralized data fusion.
%The system consists of a centralized \ac{ES} and $N$ spatially distributed \acp{SN} indexed by $\mathcal{N}\triangleq\{1,\dots,N\}$. 
  Each \ac{SN} $n \in \mathcal{N}\triangleq\{1,\dots,N\}$ contains a single \ac{AP} that primarily serves a set of single-antenna users, %denoted by $\mathcal{U}_n$, with $|\mathcal{U}_n| = U_n$, 
while also supporting radar sensing and cooperating in a bi/multistatic manner.
Each \ac{AP} is equipped with a \ac{ULA} with $M_{\mathrm a}$ antenna elements with inter-element spacing $d\leq\lambda/2$, where $\lambda$ is the carrier wavelength. %, and its boresight is characterized by an orientation angle .
The network geometry is modeled on a two-dimensional plane. Hence, the boresight angle of the \ac{ULA} is denoted by $\psi_{\mathrm n}\in[0,2\pi)$ and the location of AP $n$, denoted by $\mathbf{s}_n = (s_{n,x}, s_{n,y})$. %, and the location of user $u \in \mathcal{U}_n$ is given by $\mathbf{u}_{n,u} = (r_{n,u}, \theta_{n,u})$. 
Both $\psi_{\mathrm n}$ and $\mathbf{s}_n$ for all $n\in\mathcal{N}$ are assumed to be known at the ES, while the unknown target location $\mathbf{q} = (x_t, y_t)$ is estimated from the sensing measurements.
% The sensing service area of subnetwork $n$ is defined %separately for communication and sensing tasks as 
% as
% $\mathcal{A}_n^{(\mathrm{s})} \triangleq \{(r, \theta) \mid r_{\min}^{(\mathrm{s})} \le r \le r_{\max}^{(\mathrm{s})} \;, \theta_{\min}^{(\mathrm{s})} \le \theta \le \theta_{\max}^{(\mathrm{s})} \}$.
%where the superscript $(\mathrm{t}) \in \{\mathrm{c}, \mathrm{s}\}$ distinguishes between communication and sensing service areas.}
% A quasi-static scenario is assumed over the localization interval, i.e., APs and target remain stationary, and Doppler effects are neglected.

\subsection{Frame Structure}
The \acp{SN} operate over an \ac{OFDMA} discrete time--frequency grid % following the  principle. The OFDM waveform employs 
with $N_{\mathrm{sc}}$ subcarriers spaced in frequency by $\Delta f$, starting from the lowest frequency $f_0$. Accordingly, the total transmission bandwidth is given by $W = N_{\mathrm{sc}}\Delta f$, and the frequency of the $k$-th subcarrier is expressed as $f_k = f_0 + k\Delta f$, where $k \in\{ 0, 1, \dots, N_{\mathrm{sc}} - 1\}$.
The frame of duration $T$ is divided into $F$ time slots of equal length, where each slot contains multiple symbols and spans the system bandwidth.
% and contains $N_{sc}$ orthogonal subcarriers. 
% The 6G network ensures global synchronization, and the ES orchestrates the network-level resource allocation across all \acp{SN}. The allocation is carried out at the OFDM symbol granularity, such that each symbol is exclusively designated for either sensing or communication.%, as depicted in Fig.~\ref{fig:Fr}.}
The ES is connected to the 6G network to ensure global synchronization while also orchestrating network-level resource allocation across \acp{SN}. The allocation is carried out at the OFDM-symbol level, with each symbol assigned exclusively to either sensing or communication.
To prevent mutual interference during sensing, each sensing symbol is exclusively assigned to a single  SN,
%which transmits the probing waveform over all subcarriers, 
while the remaining \acp{SN} refrain from transmissions, and may operate in receive mode to collect echoes. %In contrast, communication slots are reused by all subnetworks simultaneously for internal downlink communication. Although different subnetworks share the communication slots non-orthogonally, each AP exploits its $M_{\mathrm a}$-element ULA to perform multiuser beamforming and spatially multiplex its internal users within the same slot/subcarriers. For simplicity, we assume $U_n\le M_{\mathrm a}$ so that AP $n$ can serve all users in $\mathcal{U}_n$ simultaneously via linear precoding.}

Without loss of generality, the localization process is performed over one or more sensing iterations within a total time budget \(T_c\), which is assumed to be sufficiently short 
%so the motion of both the \acp{SN} and the target is negligible. 
such that the target motion is negligible.
Each iteration consists of a \emph{sensing phase} and a \emph{feedback phase}. During each sensing phase, one SN acts as the probing transmitter, while the others operate in echo-reception mode or remain idle. During each feedback phase, the echo-receiving \acp{SN} report their measurements to the ES via an out-of-band control channel for centralized fusion and target localization. If further refinement is required, the ES triggers the next iteration by broadcasting the updated configuration, including the sensing symbol allocation and the designated transmitting and receiving subsets; otherwise, the localization process is terminated. Accordingly, the number of sensing iterations determines the portion of OFDM frame resources reallocated from communication to sensing.

% \begin{figure}[t]
%     \centering
%     % \includegraphics[width=1.0\linewidth]{figures/Figure_Paper_ISAC_v2CR.pdf}
%     \includegraphics[width=0.9\columnwidth]{figures/ISAC_protocol_updated_final_colors_lavender_aqua.pdf}
%     \caption{\mn{Proposed iterative OFDM sensing--communication frame. In each iteration, slots in $T_s$ are exclusively allocated to the scheduled illuminator, followed by a localization-feedback interval $T_L$; the remaining resources are shared for downlink communication.}} % \vspace{-3mm}
%     \label{fig:Fr} 
% \end{figure}

\section{Sensing framework}
%The SN APs perform multistatic sensing under the OFDMA frame structure. At each sensing iteration,
The SN APs perform multistatic sensing, where at each sensing iteration,
one SN $i\in\mathcal N$ acts as the illuminator (Tx) and a subset
$\mathcal J_i\subseteq\mathcal N\setminus\{i\}$ acts as echo receivers (Rx). During  sensing, the Tx transmits known pilots $X_i[k]$ on occupied subcarriers. After OFDM
demodulation, the received pilot sample at  Rx  $j\in\mathcal J_i$ on subcarrier $k$ and antenna $m$ is
\begin{equation}
Y_{i,j}[k,m]=X_i[k]\,H_{i,j}[k,m]+N_j[k,m],
\end{equation}
where $\mathbf H_{i,j}\in\mathbb C^{N_{\mathrm{sc}}\times M_{\mathrm a}}$ is the
bistatic echo channel coefficient and $N_j[k,m]\sim\mathcal{CN}(0,\sigma_n^2)$ is the thermal noise with power
$\sigma_n^2=k_{\mathrm B}\textrm{T}_nF_n\Delta f$, where $k_{\mathrm B}$ is the Boltzmann's constant,
$\textrm{T}_n$ is the noise temperature, and $F_n$ is the receiver noise figure.
% Neglecting Doppler within one sensing snapshot, the frequency--space echo channel across subcarriers and antennas is represented by the matrix
% $\mathbf H\in\mathbb C^{N_{\mathrm{sc}}\times M}$ whose $(k,m)$th entry collects the response on subcarrier $f_k$ and antenna $m$.
% During the sensing block, the Tx transmits known pilots $X[k]$ on the occupied subcarriers. After OFDM demodulation, the received pilot sample at Rx antenna $m$ is
% \begin{equation}
% Y[k,m] = X[k]\,H[k,m] + N[k,m],
% \end{equation}
% where $N[k,m]\sim\mathcal{CN}(0,\sigma_n^2)$ is thermal noise with $\sigma_n^2=N_0\Delta f$ (including receiver noise figure). 
Using per-tone least square (LS) estimation, we obtain
\begin{equation}
\widehat{H}_{i,j}[k,m] %\triangleq %\frac{Y_{i,j}[k,m]}{X_i[k]} 
= H_{i,j}[k,m] + W_{i,j}[k,m],
\label{eq:LS_channel_estimate}
\end{equation}
where $W_{i,j}[k,m]=N_j[k,m]/X_i[k]$ is the channel-estimation error. Under equal per-tone pilot power
$P_{\mathrm{tone}}=|X_i[k]|^2$, the entries of $\mathbf W_{i,j}$ are i.i.d.\ $\mathcal{CN}(0,\sigma_H^2)$ with
$\sigma_H^2=\sigma_n^2/P_{\mathrm{tone}}$.

The industrial environment contains widely distributed static scatterers. Therefore, the echo channel is decomposed into the intended target component $\mathbf H_{i,j}^{(\mathrm T)}$ and an aggregate clutter component $\mathbf H_{i,j}^{(\mathrm C)}$,  with the line of sight (LOS) path embedded in both, as
\begin{equation}
\mathbf H_{i,j} = \mathbf H_{i,j}^{(\mathrm T)} + \mathbf H_{i,j}^{(\mathrm C)}.
\label{eq:H_decomp}
\end{equation}

Denote the set of static clutter scatterers by $\{\mathbf c_\ell\}_{\ell\in\mathcal C}$. 
For a generic scatterer at
$\mathbf{p}=(p_x,p_y)\in\{\mathbf{q}\}\cup
\{\mathbf{c}_\ell\}_{\ell\in\mathcal{C}}$, let \(d_i(\mathbf p)\triangleq\|\mathbf p-\mathbf s_i\|\) and \(d_j(\mathbf p)\triangleq\|\mathbf p-\mathbf s_j\|\) denote the Tx--scatterer and scatterer--Rx distances. The bistatic delay is
% \begin{equation}
% d_i(\mathbf p)\triangleq\|\mathbf p-\mathbf s_i\|,\qquad
% d_j(\mathbf p)\triangleq\|\mathbf p-\mathbf s_j\|,
% \label{eq:rnggg}
% \end{equation}
% where $\|\cdot\|$ is the Euclidean norm.
% % computed after the standard polar-to-Cartesian mapping. 
% The bistatic delay is
\begin{equation}
\tau_{i,j}(\mathbf p)\triangleq\frac{d_i(\mathbf p)+d_j(\mathbf p)}{c},
\label{eq:dlyy}
\end{equation}
where $c$ is the speed of light. The AoA relative to the Rx boresight is
\begin{equation}
\vartheta_j(\mathbf{p})
\triangleq
\operatorname{wrap}_{[-\pi,\pi)}
\left(
\operatorname{atan2}
\left(
p_y-s_{j,y},
p_x-s_{j,x}
\right)
-\psi_j
\right).
\label{eq:aoa-bor}
\end{equation}
where 
% $(\cdot)_x$ and $(\cdot)_y$ denote the Cartesian components and 
$\mathrm{wrap}_{[-\pi,\pi)}(\cdot)$ maps its argument to $[-\pi,\pi)$.

For the ULA at SN $j$, with centered element index $\tilde m \triangleq m - \frac{M_{\mathrm a}-1}{2}$, 
the wideband steering response on subcarrier $k$ for an impinging angle $\vartheta$ is given by
\begin{equation}
\big[a_j(f_k,\vartheta)\big]_m \triangleq
\exp\!\Big(-j2\pi\frac{f_k}{c}\,d\sin(\vartheta)\,\tilde m\Big).
\end{equation}
With single-bounce scattering, the contribution of scatterer $\mathbf{p}$ to the $(k,m)$th channel entry is modeled as
\begin{equation}
H_{i,j}^{(\mathbf p)}[k,m]
= \gamma(\mathbf p)\, e^{-j2\pi f_k \tau_{i,j}(\mathbf p)} \,\big[a_j\!\big(f_k,\vartheta_j(\mathbf p)\big)\big]_m,
\label{eq:Hs}
\end{equation}
where $\gamma(\mathbf p)$ %captures the bistatic path loss and scattering strength. 
according to the bistatic radar equation, %it is given by 
is
$\gamma(\mathbf p)=\sqrt{G_iG_j\lambda^2\sigma(\mathbf p)/(4\pi)^3}\,\big(d_i(\mathbf p)d_j(\mathbf p)\big)^{-1}$, where $(G_i,G_j)$ denote the antenna gains and $\sigma(\mathbf p)$ is the radar 
\ac{RCS}, %which, under the Swerling-I model \cite{swerling2003probability}, 
and follows the Swerling-I model \cite{swerling2003probability} with mean $\sigma_0$.
% is exponentially distributed as
% \begin{equation}\label{eq:swerling_pdf}
% f_{\sigma}(\sigma)=\frac{1}{\sigma_0}\exp\!\Big(-\frac{\sigma}{\sigma_0}\Big),\qquad \sigma\ge 0,
% \end{equation}
% where $\sigma_0$ is the mean RCS. 
Summing \eqref{eq:Hs} over the target and clutter scatterers yields
$\mathbf H_{i,j}^{(\mathrm T)}$ and $\mathbf H_{i,j}^{(\mathrm C)}$ in \eqref{eq:H_decomp}.

\subsection{Clutter Suppression via Background Learning}
% The bistatic channel in \eqref{eq:H_decomp} is contaminated by 
The clutter term $\mathbf H^{(\mathrm C)}_{i,j}$ can obscure $\mathbf H^{(\mathrm T)}_{i,j}$. 
%, create spurious peaks, and bias the estimation of target parameters (e.g., delay/AoA) that are used downstream for localization.
Therefore, learning and suppressing $\mathbf H^{(\mathrm C)}_{i,j}$ is crucial for reliable detection, avoiding false alarms, and unbiased parameter extraction.

%The key challenge is that, although clutter is static, 
The background response is not geometry-agnostic; changing the ordered bistatic pair $(i,j)$ generally yields a different clutter channel $\mathbf H^{(\mathrm C)}_{i,j}$ due to the change in Tx/Rx locations, which necessitates repeated background learning and incurs substantial time and spectral overhead.
To address this, we exploit the fact that, in many industrial deployments, \acp{SN} operate from a finite set of feasible locations. Let $\mathcal L=\{1,\ldots,L\}$ denote the location set, and let $\pi:\mathcal N\to\mathcal L$ map each SN index to its operating location. 
% Under this finite-terminal model, the background clutter pattern is determined primarily by the \emph{ordered terminal pair} $(t,r)=(\pi(i),\pi(j))$. We therefore maintain a geometry-consistent background dictionary indexed by terminal pairs
% \begin{equation}
% \mathcal D:\mathcal L\times\mathcal L \to \mathbb C^{N_{\mathrm{sc}}\times M_{\mathrm a}},\qquad
% (t,r)\mapsto \widehat{\mathbf H}^{\mathrm{bg}}_{t,r},
% \label{eq:dict_def}
% \end{equation}
% where $\widehat{\mathbf H}^{\mathrm{bg}}_{t,r}$ denotes the learned target-free background channel for a Tx operating at terminal $t$ and an Rx operating at terminal $r$ (generally $\widehat{\mathbf H}^{\mathrm{bg}}_{t,r}\neq \widehat{\mathbf H}^{\mathrm{bg}}_{r,t}$).
Under this discrete-location model, the background clutter is determined by the ordered location pair $(t,r)=(\pi(i),\pi(j))$. Thus, we define a geometry-consistent dictionary $\mathcal{D}:\mathcal{L}\times\mathcal{L}\to\mathbb{C}^{N_{\mathrm{sc}}\times M_{\mathrm a}}$, where $(t,r)\mapsto \widehat{\mathbf{H}}^{\mathrm{bg}}_{t,r}$ denotes the learned target-free background channel, with generally $\widehat{\mathbf{H}}^{\mathrm{bg}}_{t,r}\neq \widehat{\mathbf{H}}^{\mathrm{bg}}_{r,t}$.
In practice, $\mathcal D$ can be stored at the \ac{ES} and periodically distributed to \acp{SN}; each receiver then performs clutter suppression by simple dictionary lookup using its own location and the active Tx location.

To populate the dictionary $\mathcal D$,
a preprocessing stage is performed prior to 
%the execution of any localization task by the ES.
any localization task by the ES.
%in the preprocessing stage, let
Let $N_{\mathrm{BG}}$ denote the number of target-free symbols used to learn one dictionary entry. Calibration is performed sequentially over feasible transmit locations; for each $t\in\mathcal L$, a node operating at location $t$ transmits pilots once while receivers at all other locations $r\in\mathcal L\setminus\{t\}$ collect target-free observations simultaneously. Hence, the set of entries $\{\widehat{\mathbf H}^{\mathrm{bg}}_{t,r}\}_{r\neq t}$ is obtained. 
% within a single calibration block.
Denoting the resulting matrices by $\{\widehat{\mathbf H}^{\mathrm{bg},(n)}_{t,r}\}_{n=1}^{N_{\mathrm{BG}}}$, %and let $\mathrm{med}_{n}\{\cdot\}$ denote the entrywise %sample 
%median over $n=1,\ldots,N_{\mathrm{BG}}$. For a fixed ordered terminal pair $(t,r)$ with $r\neq t$, %since the clutter is quasi-static over the learning window, 
we compute a robust background estimate via the complex median %$\mathrm{med}_{n}\{\cdot\}$
\begin{equation}
\widehat{\mathbf H}^{\mathrm{bg}}_{t,r}
\triangleq
\mathrm{med}_{n}\!\Big(\Re\{\widehat{\mathbf H}^{\mathrm{bg},(n)}_{t,r}\}\Big)
+j\,\mathrm{med}_{n}\!\Big(\Im\{\widehat{\mathbf H}^{\mathrm{bg},(n)}_{t,r}\}\Big),
\label{eq:bg_median_terminals}
\end{equation}
and store it as the dictionary entry $\mathcal D(t,r)\triangleq \widehat{\mathbf H}^{\mathrm{bg}}_{t,r}$. The resulting dictionary can subsequently be employed for clutter suppression across localization tasks and updated over time to reflect changes in the clutter environment.
% During the actual sensing snapshot, the receiver first obtains $\widehat{\mathbf H}_{i,j}$ via \eqref{eq:LS_channel_estimate} and then retrieves the corresponding geometry-consistent background from the dictionary
% \begin{equation}
% \widehat{\mathbf H}^{\mathrm{bg}}_{\pi(i),\pi(j)} \triangleq \mathcal D\!\big(\pi(i),\pi(j)\big).
% \label{eq:dict_lookup}
% \end{equation}
% The clutter-suppressed residual channel is then

During sensing, receiver \(j\) first obtains \(\widehat{\mathbf H}_{i,j}\) from the LS estimate and then retrieves the geometry-consistent background \(\widehat{\mathbf H}^{\mathrm{bg}}_{\pi(i),\pi(j)}=\mathcal D(\pi(i),\pi(j))\). The clutter-suppressed residual channel is therefore
\begin{align}
\widetilde{\mathbf H}_{i,j}
&\triangleq \widehat{\mathbf H}_{i,j}-\widehat{\mathbf H}^{\mathrm{bg}}_{\pi(i),\pi(j)} \nonumber\\
&= \mathbf H^{(\mathrm T)}_{i,j}
+ \underbrace{\big(\mathbf H^{(\mathrm C)}_{i,j}-\widehat{\mathbf H}^{\mathrm{bg}}_{\pi(i),\pi(j)}\big)}_{\text{residual clutter}}
+ \widetilde{\mathbf W}_{i,j},
\label{eq:residual}
\end{align}
where $\widetilde{\mathbf W}_{i,j}$ aggregates the sensing-snapshot LS error and the median-attenuated calibration noise in the dictionary entry. 
%If the background is accurately learned, the residual clutter term is small and $\widetilde{\mathbf H}_{i,j}$ is dominated by the target echo plus estimation noise, enabling reliable extraction of target parameters.

\subsection{Range--Angle Power Map, Detection, and Refinement}

Given the background-suppressed residual channel $\widetilde{\mathbf H}_{i,j}\in\mathbb{C}^{N_{\mathrm{sc}}\times M_{\mathrm a}}$, we construct a range--angle power map that highlights target-induced peaks.
%and serves as the input to the subsequent 2D-CFAR detector.
Let $\widetilde{\mathbf h}_{i,j}[k]\triangleq[\widetilde{\mathbf H}_{i,j}]_{k,:}^{\mathsf T}\in\mathbb{C}^{M_{\mathrm a}\times 1}$ denote the spatial snapshot on subcarrier $k$. The receiver performs beam sweeping over a predefined grid of candidate angles $\Theta=\{\vartheta_g\}_{g=1}^{N_\vartheta}$.  
%For each $\vartheta_g$, we apply wideband matched receive combining across the ULA using the steering vector $\mathbf a_j(f_k,\vartheta_g)\in\mathbb{C}^{M_{\mathrm a}\times 1}$, yielding the beamformed frequency response
For each $\vartheta_g$, wideband matched receive combining across the ULA using the steering vector $\mathbf a_j(f_k,\vartheta_g)\in\mathbb{C}^{M_{\mathrm a}\times 1}$, yields
\begin{equation}
z_{i,j}[k;\vartheta_g]
\triangleq
\frac{1}{\sqrt{M_{\mathrm a}}}\,\mathbf a_j^\dagger(f_k,\vartheta_g)\,\widetilde{\mathbf h}_{i,j}[k],
\label{eq:wb_bf_freq}
\end{equation}
% To reduce delay sidelobes, a Hann window is applied as a frequency-domain taper $w[k]$ across subcarriers. The angle-conditioned delay profile is
% then obtained by 
% %delay compression across $k$ as
% \begin{equation}
% Z_{i,j}[\ell;\vartheta_g]
% \triangleq
% \mathrm{IFFT}_{k}\!\left\{\,w[k]\;z_{i,j}[k;\vartheta_g]\right\},\qquad
% \ell=0,\ldots,N_{\mathrm{sc}}-1.
% \label{eq:wb_bf_range}
% \end{equation}
% Because the subcarriers are uniformly spaced, \eqref{eq:wb_bf_range} yields a sampled delay response with bin spacing
% $\Delta\tau=1/(N_{\mathrm{sc}}\Delta f)$, i.e., $\tau_\ell=\ell\,\Delta\tau$, and the associated bistatic range is $\rho_\ell=c\,\tau_\ell$.
% Finally, the single-shot wideband angle--range power map 
% is 
% %defined entrywise as
% \begin{equation}
% P_{i,j}[g,\ell]\triangleq \big|Z_{i,j}[\ell;\vartheta_g]\big|^2,
% \label{eq:angle_range_map}
% \end{equation}
% and collecting \eqref{eq:angle_range_map} over $g=1,\ldots,N_\vartheta$ and $\ell=0,\ldots,N_{\mathrm{sc}}-1$ yields
% $\mathbf P_{i,j}\in\mathbb{R}_+^{N_\vartheta\times N_{\mathrm{sc}}}$. The map $\mathbf P_{i,j}$ is subsequently processed by a 2-D CFAR
% detector to produce candidate angle--range bins $(\vartheta_g,\ell_0)$, which provide initial AoA/delay hypotheses for downstream 
% localization.
where \((\cdot)^\dagger\) is the Hermitian transpose. The angle-conditioned delay profile is then obtained by delay compression across subcarriers, i.e., \(Z_{i,j}[\ell;\vartheta_g]=\mathrm{IFFT}_{k}\{z_{i,j}[k;\vartheta_g]\}\) for \(\ell=0,\ldots,N_{\mathrm{sc}}-1\). This yields a sampled delay response with bin spacing \(\Delta\tau=1/(N_{\mathrm{sc}}\Delta f)\), corresponding to bistatic range \(\rho_\ell=c\ell\Delta\tau\). The resulting %single-shot 
range--angle power map is
\begin{equation}
P_{i,j}[g,\ell]\triangleq \big|Z_{i,j}[\ell;\vartheta_g]\big|^2.
\end{equation}
Collecting \(P_{i,j}[g,\ell]\) over all angle and delay bins yields $\mathbf P_{i,j}\in\mathbb{R}_+^{N_\vartheta\times N_{\mathrm{sc}}}$. %,whose dominant peaks provide coarse \ac{AoA}--delay hypotheses for subsequent processing.
For each sensing link, the range--angle power map \(\mathbf P_{i,j}\) is processed using an edge-aware 2-D CA-CFAR detector~\cite{rohling2007radar} to identify candidate angle--range bins and extract coarse hypotheses for subsequent localization. The resulting detections are further pruned using the \ac{NMS} technique to avoid multiple detections from the same peak region, yielding the candidate set \({\mathcal K}_{i,j}\).
For each candidate \((g_0,\ell_0)\in{\mathcal K}_{i,j}\) with coarse angle \(\vartheta_{g_0}\) and coarse delay \(\tau_0=\ell_0\Delta\tau\), 
%the AoA and ToA are refined from the complex antenna-domain observations via MUSIC-based angular super-resolution~\cite{xhafa2025experimental} 
%and phase-slope delay refinement~[REF], 
%while the delay is refined from the linear phase trend across subcarriers,
the AoA is refined via MUSIC-based angular super-resolution~\cite{xhafa2025experimental}, while the ToA is refined from the linear phase trend across subcarriers,
yielding a refined AoA--ToA pair \((\widehat{\vartheta}_{i,j},\widehat{\tau}_{i,j})\).

\section{Data Fusion}
For each active sensing link $(i,j)$, receiver $j$ reports the \ac{ToA} estimate $\widehat{\tau}_{i,j}$, the \ac{AoA} estimate $\widehat{\vartheta}_{i,j}$, and the corresponding \ac{eCRLB}s as reliability indicators to the ES.
%that quantifies the reliability of the measurement.
These \ac{eCRLB}s are computed from the detected-cell power and a robust estimate of the noise floor on the range--angle map.
Specifically, let $P^{\mathrm{det}}_{i,j}\triangleq P_{i,j}[g_0,\ell_0]$ denote the power of the selected detection, and let $\widehat{P}^{\mathrm{n}}_{i,j}\triangleq \mathrm{median}\{P_{i,j}[g,\ell]\}_{g,\ell}$ denote a robust estimate of the noise floor. Based on these quantities, we define the SNR proxy $\widehat{\gamma}_{i,j}\triangleq P^{\mathrm{det}}_{i,j}/(\widehat{P}^{\mathrm n}_{i,j}+\epsilon)$ and the corresponding \ac{ToA} %standard deviation
variance~\cite{gezici2009position} is approximated as $\sigma^2_{\tau,i,j}\triangleq \big(2\pi W_{\mathrm{rms}}\sqrt{2\widehat{\gamma}_{i,j}}\big)^{-2}$ with
$W_{\mathrm{rms}}\triangleq \sqrt{\frac{1}{N_{\mathrm{sc}}}\sum_{k} f_k^2}$. Likewise, the \ac{AoA} variance~\cite{xhafa2025experimental} is approximated as  
\mbox{$\sigma^2_{\vartheta,i,j}\triangleq 6\left({N_{\mathrm{s}}\,\widehat{\gamma}_{i,j}\big(2\pi d/\lambda\big)^2 M_{\mathrm a}(M_{\mathrm a}^2-1)\cos^2(\widehat{\vartheta}_{i,j})}\right)^{-1}$}.
The \ac{ES} then estimates the target location $\mathbf q$ %by solving the nonlinear WLS problem
via  nonlinear WLS %problem
\begin{equation}\label{eq:P1_fusion}
\begin{aligned}
\textbf{P1:}\quad
%\min_{\mathbf q\in\mathcal A}\;
\min_{\mathbf q}\;
&\sum_{j\in\mathcal J_i^{\tau}}
\frac{\big(\widehat{\tau}_{i,j}-\tau_{i,j}(\mathbf q)\big)^2}{\sigma_{\tau,i,j}^2}
\\
&\quad+
\sum_{j\in\mathcal J_i^{\vartheta}}
\frac{\big(
\mathrm{wrap}_{[-\pi,\pi)}(\widehat{\vartheta}_{i,j}-\vartheta_{j}(\mathbf q))
\big)^2}{\sigma_{\vartheta,i,j}^2},
\end{aligned}
\end{equation}
where $\mathcal J_i^{\tau}$ and $\mathcal J_i^{\vartheta}$ denote the sets of receivers that provide admitted \ac{ToA} and \ac{AoA} measurements, respectively. Problem~\textbf{P1} is a nonlinear WLS problem that can be efficiently solved by Gauss--Newton iterations~\cite{nozari2025toward}.
%$\tau_{i,j}(\mathbf q)$ and $\vartheta_j(\mathbf q)$ follow directly from \eqref{eq:dlyy} and \eqref{eq:aoa-bor}, respectively, $\mathcal A$ is the operational area, 
% $\mathcal J_i^{\tau}$ denotes the set of receivers providing valid delay measurements, and $\mathcal J_i^{\vartheta}$ denotes the set of receivers whose AoA measurements are admitted. 

\section{Iterative Adaptive Sensing, Localization, and Node-Selection Planning}
The \ac{ES} executes 
%iterative PEB-Based Sensing Reconfiguration(PEB--SR) scheme for cooperative ISAC localization. 
the sensing--localization pipeline iteratively. %In the initial sensing block, it randomly selects a transmitter $i^{(1)} \in \mathcal N$ and a receiver subset $\mathcal J^{(1)} \subseteq \mathcal N \setminus \{i^{(1)}\}$ with $|\mathcal J^{(1)}|=M$. It then fuses the reported measurements by solving Problem~\textbf{P1}, thereby obtaining an initial target-location estimate $\widehat{\mathbf q}^{(1)}$.
At the initial iteration, no estimate of the target location is available. Therefore, selecting the Tx--Rx subset based on a random assumed target point can bias the algorithm toward a weak initial geometry and degrade the subsequent updates. To mitigate this, we adopt a sampling-based FoV-aware initialization (SFI). Specifically, \(N_{sp}\) sample points are uniformly drawn over the target search area, and the initial receiver subset is chosen to maximize their FoV coverage
\begin{equation}
\mathcal J^{(1)\star}
\coloneqq
\arg\max_{\substack{\mathcal J \subseteq \mathcal N\\ |\mathcal J|=N_r}}
\sum_{s=1}^{N_{sp}}
\mathbf 1\!\left(\exists j\in\mathcal J:\, I_{j,s}=1\right),
\label{eq:init_rx}
\end{equation}
where \(I_{j,s}=1\) if sample point \(s\)
lies inside the FoV of receiver \(j\) 
%lies in receiver \(j\)'s FoV,
and \(0\) otherwise. Thus, the initial receiver subset is chosen to 
maximize geometric visibility over the target search area.
%provide the broadest geometric visibility over the target search area.
Given \(\mathcal J^{(1)\star}\), the initial transmitter is selected to minimize the average empirical PEB over the same sample points as
\begin{IEEEeqnarray}{c}
i^{(1)\star}
\coloneqq
\arg\min_{i\in\mathcal N\setminus \mathcal J^{(1)\star}}
\frac{1}{|\mathcal X_{\mathrm{use}}|}
\sum_{\mathbf q\in\mathcal X_{\mathrm{use}}}
\mathrm{PEB}_{\mathrm{}}\!\left(\mathbf q;\, i,\mathcal J^{(1)\star}\right),\IEEEeqnarraynumspace
\label{eq:init_tx}
\end{IEEEeqnarray}
where \(\mathcal X_{\mathrm{use}}\) denotes the sampled area points,
% used in the initialization stage
and \(\mathrm{PEB}_{\mathrm{}}(\cdot)\) denotes the empirical localization-quality metric, whose explicit expression is given later in \eqref{eq:peb_score}. 
% This initialization improves the likelihood of starting from an informative sensing geometry. 
The resulting configuration \(\bigl(i^{(1)\star},\mathcal J^{(1)\star}\bigr)\) is then used to obtain the first target-location estimate \(\widehat{\mathbf q}^{(1)}\) by solving Problem~\textbf{P1}.

At iteration $j > 1$, the ES uses the current estimate $\widehat{\mathbf q}^{(j)}$ to select the next multistatic sensing configuration. Specifically, it selects the transmitter $i^{(j+1)} \in \mathcal N$ and the receiver subset $\mathcal J^{(j+1)} \subseteq \mathcal N \setminus \{i^{(j+1)}\}$, with $|\mathcal J^{(j+1)}|=N_r$, that yield the smallest predicted PEB, defined later in Problem~\textbf{P2}. The selected nodes then perform new sensing measurements, and report them to the ES for data fusion and refining the target-location estimate to $\widehat{\mathbf q}^{(j+1)}$ by solving Problem~\textbf{P1}.

This iterative procedure continues until the sensing reconfiguration remains unchanged over two consecutive iterations, that is, when $i^{(j+1)} = i^{(j)}$ and $\mathcal J^{(j+1)} = \mathcal J^{(j)}$. The overall procedure of PEB-based sensing reconfiguration (PEB-SR) is summarized in Algorithm~1.

\begin{algorithm}[t]
\footnotesize
\caption{PEB-Based Sensing Reconfiguration (PEB-SR) for Cooperative Localization}
\label{alg:peb_isac}
\LinesNumbered
\KwIn{SN locations and boresights $\{(\mathbf{s}_n,\psi_n)\}_{n=1}^{N}$, 
Rx set cardinality $|\mathcal{J}|=N_r$, max. no. of reconfigurations $J_{\max}$, 
max. no. of WLS iterations $L_{\max}$, threshold $\varepsilon$.}
\KwOut{Estimated location $\hat{\mathbf{q}}^\star$, selected configuration $(i^\star,\mathcal{J}^\star)$.}
\BlankLine

\textbf{Initialize:}\ 
Select an initial sensing configuration $(i^{(1)},\mathcal{J}^{(1)})$ via~~\eqref{eq:init_rx} and~\eqref{eq:init_tx};\ 
initialize $\hat{\mathbf q}$;\ set $j\gets 1$\;

\While{$j \le J_{\max}$}{%

  \textbf{Sensing:}\ Activate Tx $i^{(j)}$ and receivers $\mathcal{J}^{(j)}$\;
  \ForEach{$r\in\mathcal{J}^{(j)}$}{%
    Receiver $r$ estimates ToA and uncertainty $\big(\hat{\tau}_{i^{(j)},r},\,\sigma_{\tau,i^{(j)},r}^{2}\big)$\;
    Receiver $r$ estimates AoA and uncertainty $\big(\hat{\vartheta}_{i^{(j)},r},\,\sigma_{\vartheta,i^{(j)},r}^{2}\big)$\;
    Send measured values to the ES\;
  }

  \BlankLine
  \textbf{Data fusion (P1 at ES):}\ Initialize $\mathbf{q}^{(0)} \gets \hat{\mathbf q}$\;
   \For{$\ell \gets 1$ \KwTo $L_{\max}$}{%
    Update $\mathbf q^{(\ell)}$ by one Gauss--Newton WLS step in \textbf{P1}\;
    \If{$\|\mathbf{q}^{(\ell)}-\mathbf{q}^{(\ell-1)}\|\le \varepsilon$}{%
      \textbf{break}\;
    }
  }
  Set $\hat{\mathbf q} \gets \mathbf{q}^{(\ell)}$\;

  \BlankLine
  $j \gets j+1$\;
  \textbf{PEB-based configuration selection (P2 at ES):}\ 
  Evaluate $\mathrm{PEB}(i,\mathcal{J})$ for all $i\in\mathcal{N}$ and
  $\mathcal{J}\subseteq\mathcal{N}\setminus\{i\}$ using $\hat{\mathbf q}$ via~\eqref{eq:peb_score}\;
  $(i^{(j)},\mathcal{J}^{(j)}) \gets \arg\min \mathrm{PEB}(i,\mathcal{J})$\;

  \If{$(i^{(j)},\mathcal{J}^{(j)})=(i^{(j-1)},\mathcal{J}^{(j-1)})$}{%
    \textbf{break}\;
  }
}
\Return $(i^\star,\mathcal{J}^\star)\gets(i^{(j-1)},\mathcal{J}^{(j-1)})\;,$
$\hat{\mathbf{q}}^\star \gets \hat{\mathbf{q}}$\;
\end{algorithm}

To schedule the next sensing iteration, the ES evaluates all possible sensing configurations using the current estimate $\widehat{\mathbf q}^{(j)}$. In fact, $\widehat{\mathbf q}^{(j)}$ serves as a proxy target location for predicting the geometric sensitivity and measurement reliability of candidate links that were inactive in the previous iteration. For simplicity, we denote $\widehat{\mathbf q}\triangleq \widehat{\mathbf q}^{(j)}$.
For a candidate bistatic link $(i,j)$, the gradient of the bistatic delay in \eqref{eq:dlyy}, evaluated at $\widehat{\mathbf q}$, is
\vspace{-0.7em}
\begin{equation}
\mathbf g_{\tau,i,j}(\widehat{\mathbf q})
=
\nabla_{\mathbf q}\tau_{i,j}(\mathbf q)\big|_{\mathbf q=\widehat{\mathbf q}}
=
\frac{1}{c}
\left(
\frac{\widehat{\mathbf q}-\mathbf s_i}{d_i(\widehat{\mathbf q})}
+
\frac{\widehat{\mathbf q}-\mathbf s_j}{d_j(\widehat{\mathbf q})}
\right)
% \in\mathbb{R}^{2\times 1}
.
\label{eq:gtau}
\end{equation}
Likewise, the gradient of the AoA in \eqref{eq:aoa-bor}, evaluated at $\widehat{\mathbf q}$, is
\begin{equation}
\mathbf g_{\vartheta,j}(\widehat{\mathbf q})
=
\nabla_{\mathbf q}\vartheta_j(\mathbf q)\big|_{\mathbf q=\widehat{\mathbf q}}
=
\frac{1}{\|\widehat{\mathbf q}-\mathbf s_j\|^2}
\begin{bmatrix}
-(\widehat q_y-s_{j,y})\\
\ \ (\widehat q_x-s_{j,x})
\end{bmatrix}
% \in\mathbb{R}^{2\times 1}
.
\label{eq:gtheta}
\end{equation}
% Using the bistatic radar-equation scaling, the per-tone channel-amplitude proxy at $\widehat{\mathbf q}$ is modeled as
% \begin{equation}
% \alpha_{i,j}(\widehat{\mathbf q})
% =
% \frac{\lambda}{(4\pi)^{3/2}}\,
% \frac{\sqrt{G_i G_j\,\sigma(\widehat{\mathbf q})}}{d_i(\widehat{\mathbf q})\,d_j(\widehat{\mathbf q})}.
% \label{eq:alpha_pred}
% \end{equation}
% A proxy per-tone SNR is then defined as $\widetilde{\gamma}_{i,j}\triangleq |\alpha_{i,j}(\widehat{\mathbf q})|^2/\sigma_H^2$.
Using the bistatic radar-equation scaling, the per-tone channel-amplitude proxy at $\widehat{\mathbf q}$ is modeled as $\alpha_{i,j}(\widehat{\mathbf q}) = \sqrt{G_i G_j \lambda^2 \sigma(\widehat{\mathbf q})/(4\pi)^3}\,\big(d_i(\widehat{\mathbf q})d_j(\widehat{\mathbf q})\big)^{-1}$, and the proxy per-tone SNR is then defined as $\widetilde{\gamma}_{i,j}\triangleq \alpha_{i,j}^2(\widehat{\mathbf q})/\sigma_H^2$.
Given $\widetilde{\gamma}_{i,j}$, the predicted ToA and AoA measurement variances, denoted by $\sigma_{\tau,i,j}^2(\widehat{\mathbf q})$ and $\sigma_{\vartheta,i,j}^2(\widehat{\mathbf q})$, respectively, are obtained using the same models as in Problem~\textbf{P1}. Accordingly, the predicted Fisher-information contribution of link $(i,j)$ is
% at the proxy location $\widehat{\mathbf q}$ is

\begin{equation}
\begin{aligned}
\mathbf J_{i,j}(\widehat{\mathbf q})
&\triangleq
\frac{1}{\sigma_{\tau,i,j}^2(\widehat{\mathbf q})}\,
\mathbf g_{\tau,i,j}(\widehat{\mathbf q})
\mathbf g_{\tau,i,j}^{\mathsf T}(\widehat{\mathbf q}) \\
&\quad+
\frac{1}{\sigma_{\vartheta,i,j}^2(\widehat{\mathbf q})}\,
\mathbf g_{\vartheta,j}(\widehat{\mathbf q})
\mathbf g_{\vartheta,j}^{\mathsf T}(\widehat{\mathbf q}).
\end{aligned}
\label{eq:link_fim_pred}
\end{equation}
% By the additive property of Fisher information, the total predicted information for a candidate configuration $(i,\mathcal J)$ is
% \begin{equation}
% \mathbf J_{i,\mathcal J}(\widehat{\mathbf q})
% \triangleq
% \sum_{j\in\mathcal J}\mathbf J_{i,j}(\widehat{\mathbf q}).
% \label{eq:subset_fim_pred}
% \end{equation}
By the additive property of Fisher information, the total predicted information for a candidate configuration \((i,\mathcal J)\) is \(\mathbf J_{i,\mathcal J}(\widehat{\mathbf q}) \triangleq \sum_{j\in\mathcal J}\mathbf J_{i,j}(\widehat{\mathbf q})\).
Accordingly, the regularized information matrix is formed as
$\mathbf A_{i,\mathcal J}\triangleq \mathbf J_{i,\mathcal J}(\widehat{\mathbf q})+\mu_{\mathrm{reg}}\mathbf I_2$,
where $\mu_{\mathrm{reg}}>0$ is a small regularization constant, and the candidate configuration is scored by the empirical \ac{PEB},
\begin{equation}
\mathrm{PEB}(i,\mathcal J)
\triangleq
\sqrt{\mathrm{tr}\!\left(\mathbf A_{i,\mathcal J}^{-1}\right)}.
\label{eq:peb_score}
\end{equation}
The next sensing configuration is then selected by solving ~\textbf{P2},
\begin{equation}
\begin{aligned}
\textbf{P2:}\quad
(i^\star,\mathcal J^\star)
\coloneqq
\arg\min_{\substack{i\in\mathcal N,\,\mathcal J\subseteq \mathcal N\setminus\{i\}\\|\mathcal J|=N_r}}
\mathrm{PEB}(i,\mathcal J).
\end{aligned}
\label{eq:P2_selection}
\end{equation}
The next sensing iteration is scheduled using $i^\star$ and $\mathcal J^\star$. Problem~\textbf{P2} is combinatorial; however, exhaustive search remains computationally feasible for moderate values of $|\mathcal N|$ and  $N_r$.

\section{Numerical Results and Discussion} \label{Numerical}
% In this section, we evaluate the performance of the proposed algorithm. %The main simulation parameters are summarized in  Table~\ref{tab:sim_params}. 
% We consider an industrial scenario where a target and \(N=40\) subnetworks are independently and uniformly distributed over a \(200 \times 200\,\mathrm{m}^2\) area. The carrier frequency is \(10\,\mathrm{GHz}\), and the frame structure follows the 5G NR specification with numerology~0.  Specifically, each frame has a duration of \(T = 10\,\mathrm{ms}\) and is divided into \(F = 10\) slots, where each slot lasts \(1\,\mathrm{ms}\) and consists of 14 OFDM symbols.  The total system bandwidth is \(W = 100\,\mathrm{MHz}\). Unless otherwise stated, the sensing duration per iteration is set to one OFDM symbol.
%In this section, we evaluate the proposed scheme in an \ac{IIoT} scenario where one target and \(N=40\) SNs are uniformly distributed over a \(200 \times 200\,\mathrm{m}^2\) area.
In this section, we evaluate the proposed scheme in an \ac{IIoT} scenario where one target, \(N=40\) SNs, and \(N_{\mathrm{cl}}=30\) dedicated clutter scatterers are uniformly distributed over a \(200 \times 200\,\mathrm{m}^2\) area, with the SNs also treated  as additional clutter sources.
The carrier frequency is \(10\,\mathrm{GHz}\), the transmit power is \(p_n=23\,\mathrm{dBm}\), the mean target/clutter RCS is \(\sigma_0=0\,\mathrm{dBsm}\), and the transmit and receive antenna gains are \(G_t=G_r=0\,\mathrm{dBi}\), with boresight angle $\psi_n \sim \mathcal{U}[0,2\pi)$. The frame structure follows 5G NR numerology~0, with frame duration \(T=10\,\mathrm{ms}\), \(F=10\) slots per frame, and 14 OFDM symbols per slot. The bandwidth is \(W=100\,\mathrm{MHz}\). %Unless otherwise stated, each sensing iteration occupies one OFDM symbol. 
The number of background-learning symbols is set to \(N_{\mathrm{BG}}=70\), and the noise power spectral density is \(-174\,\mathrm{dBm/Hz}\).

We compare the proposed scheme with three reference variants, three benchmark schemes, and two theoretical bounds. The reference variants are the proposed scheme with no clutter suppression (PS--NCS), assuming clutter-free conditions (PS--CF), and omitting SFI (PS--NSFI). The benchmarks %include one iterative and two single-shot schemes. 
and the theoretical bounds are:
\begin{itemize}
    % \item \textbf{Random Subset Averaging (RSA):} an iterative random subset averaging (RSA) scheme. The first subset is selected using SFI, while the remaining subsets are chosen randomly. The final estimate is obtained by averaging the intermediate estimates.
    \item \textbf{Random Subset Averaging (RSA):} an iterative scheme in which the first subset is selected using SFI and the next subsets are chosen randomly. The final estimate is obtained by averaging the intermediate estimates.
    \item \textbf{Fixed Initialized Subset (FIS):} a single-shot scheme using the subset selected by SFI. %with \(J_{\max}\) sensing symbols.
    \item \textbf{Oracle Best Subset (OBS):} a single-shot scheme that assumes perfect knowledge of the true target location and selects the optimal sensing subset from \eqref{eq:P2_selection}.

    \item \textbf{OBS-PEB:} a single-shot theoretical lower bound, corresponding to the OBS, obtained via~\eqref{eq:peb_score}.
    \item \textbf{SFI-PEB:} a single-shot theoretical lower bound, corresponding to the subset selected by SFI, obtained via~\eqref{eq:peb_score}.
\end{itemize}
 For a fair comparison, the proposed method and the RSA use one sensing symbol per iteration over \(J_{\max}\) iterations, while each single-shot benchmark is assigned the same total budget of \(J_{\max}\) symbols, %all used 
in a single sensing iteration.

\begin{figure}[!t]
    \centering
    \includegraphics[width=0.95\columnwidth]{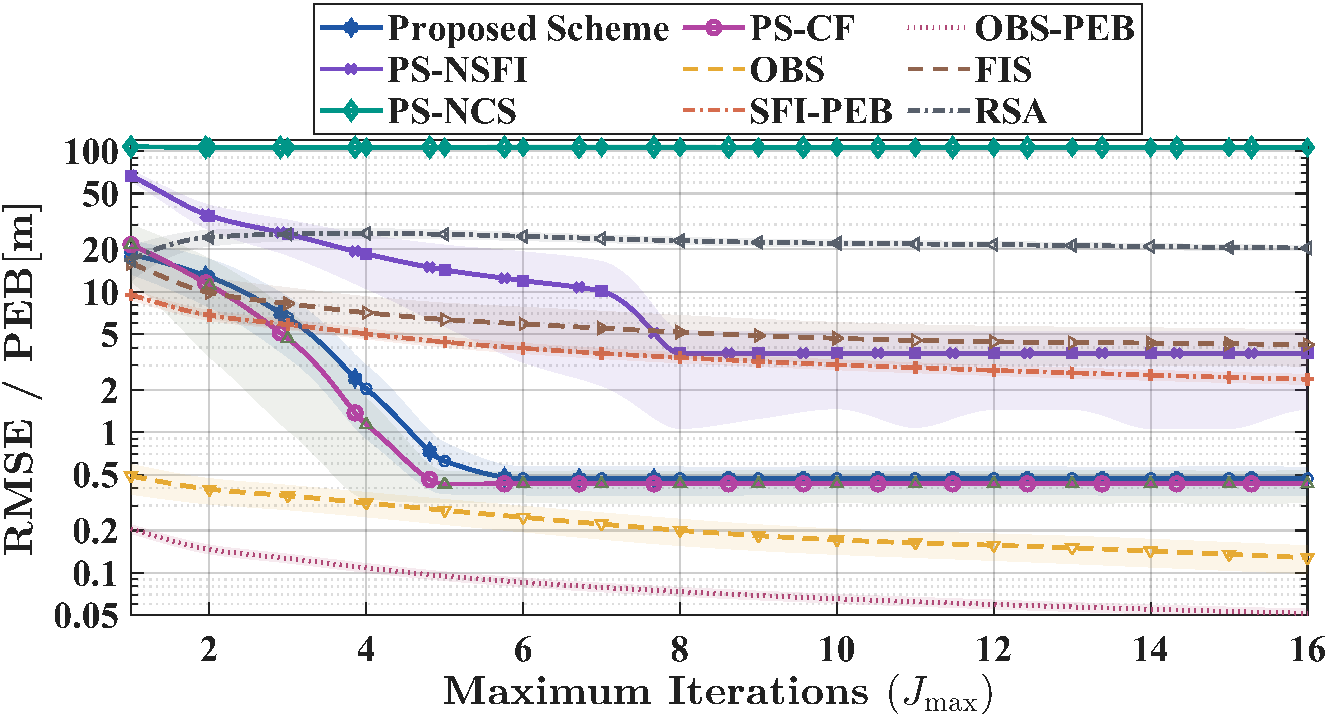}\vspace{-0.3em}
     % \caption{Performance comparison of the proposed adaptive scheme, its variants (NSI, NCS, and CF), and benchmark schemes.}
     \caption{Accuracy of the proposed method against benchmark schemes.}
    \label{fig:deployment}
\end{figure}

Fig.~\ref{fig:deployment} shows the empirical RMSE and corresponding PEB versus \(J_{\max}\) for the bistatic setup with \(M_a=32\) antennas. The shaded regions denote the \(90\%\) confidence intervals. 
The proposed scheme rapidly reduces the localization RMSE from about \(20\)~m at \(J_{\max}=1\) to about \(0.45\)~m at \(J_{\max}=6\), after which the performance saturates. The  PS--CF variant follows a similar trend, with only slightly lower RMSE during the transient regime, indicating that the proposed clutter-suppression method effectively recovers clutter-free accuracy. In contrast, the  PS--NCS variant remains close to \(100\)~m over the entire range of \(J_{\max}\), confirming that clutter suppression is essential. The PS--NSFI variant converges more slowly and to a much higher error floor, decreasing from about \(67\)~m to only \(3.6\)~m, with a wider \(90\%\) confidence interval that indicates greater sensitivity to the initial SN choice. This trend highlights the importance of SFI when no prior information about the target location is yet available.
% Among the benchmark schemes, OBS serves as an oracle-aided reference for the best achievable sensing performance within the current framework. Its RMSE decreases from about $0.45$~m at $J_{\max}=1$ to approximately $0.13$~m at $J_{\max}=16$, while the corresponding OBS-PEB, as expected, provides a lower bound and decreases from about $0.2$~m to nearly $0.05$~m over the same range. Since OBS assumes access to the true target location for subset selection, it should be interpreted as an optimistic empirical lower bound rather than a practically realizable scheme. At the convergence point of the proposed scheme, i.e., around $J_{\max}=6$, OBS attains an RMSE of approximately $0.25$~m, indicating that the proposed method operates relatively close to the oracle benchmark despite relying only on progressively refined target estimates. The remaining performance gap is mainly attributable to oracle-aided subset selection and the use of the full sensing budget in a single sensing shot.
Among the benchmark schemes, OBS achieves the best empirical localization performance. 
% . Its RMSE decreases from about \(0.45\)~m at \(J_{\max}=1\) to \(0.13\)~m at \(J_{\max}=16\), while the corresponding OBS-PEB drops from about \(0.2\)~m to nearly \(0.05\)~m. 
However, since it assumes perfect knowledge of the target location for subset selection, it should be regarded as an optimistic lower reference rather than a realizable scheme.
% Since OBS assumes perfect knowledge of the true target location for subset selection, it should be viewed as an optimistic lower reference rather than a realizable scheme. 
At 
%the convergence point of the proposed method, 
\(J_{\max}=6\), OBS achieves an RMSE of about \(0.25\)~m, only about \(0.2\)~m below the proposed method; this remaining gap is mainly %attributable 
due
to oracle-aided subset selection and one-shot use of the full sensing budget.
%the use of the full sensing budget in a single sensing shot.
The FIS improves monotonically with \(J_{\max}\), with its RMSE decreasing from about \(20\)~m to \(4.2\)~m, while the corresponding SFI-PEB stays lower and decreases from about \(10\)~m to \(2.4\)~m. 
%By contrast, RSA exhibits non-monotonic behavior. Although it starts at about \(20\)~m thanks to the proposed SFI-based initialization, the subsequent random subset updates occasionally select geometrically weak SNs, leading to an increase in the RMSE to about \(26\)~m over the next few iterations before averaging reduces it to about \(20\)~m at \(J_{\max}=16\). 
By contrast, RSA exhibits non-monotonic behavior. Although it initially benefits from the proposed SFI scheme, subsequent random subset updates increase the RMSE over the next few iterations before averaging reduces it again.
% Overall
% %, Fig.~\ref{fig:deployment} 
% %shows that the proposed scheme provides the best empirical accuracy among the realizable benchmarks and is closest to OBS.
% %shows that
% the proposed scheme achieves the best accuracy among the realizable benchmarks and is closest to OBS.
Overall, the combination of SFI and PEB-based node refinement yields a nearly two orders-of-magnitude RMSE reduction in six iterations, the lowest error among realizable benchmarks, and the smallest gap to OBS.

Table~\ref{tab:percentile_final_error} reports the percentile localization error at convergence for different \((M_a,N_r)\) configurations. Increasing \(M_a\) consistently improves performance across all percentiles, reflecting the benefit of larger arrays for more accurate AoA estimation. By contrast, the impact of increasing \(N_r\) is non-monotonic and percentile-dependent: it generally improves the upper tail of the error distribution, e.g., the \(95\)th percentile, by reducing large-error events through multistatic diversity, but does not uniformly improve the lower and central percentiles, since additional receivers may provide weak or geometrically unfavorable observations. This suggests that adaptive \(N_r\) selection based on sensing geometry and measurement quality is a promising direction for future work.

\begin{table}[t]
% \caption{Percentiles of RMSE localization error at convergence for different $(N_{\mathrm{r}}, M)$ configurations}
\caption{Percentile localization error for different $(M_a,N_{\mathrm{r}})$ configurations.}\vspace{-0.7em}
\label{tab:percentile_final_error}
\centering
\footnotesize
\setlength{\tabcolsep}{3.1pt}
\renewcommand{\arraystretch}{1.1}
\begin{tabular}{@{}c c c c c c c c@{}}
\hline
Percentile (\%) & \multicolumn{3}{c}{$M_a=4$} & &\multicolumn{3}{c}{$M_a=16$} \\
\cline{2-4}\cline{6-8}
& $N_r=1$ & $N_r=2$ & $N_r=3$ & &$N_r=1$ & $N_r=2$ & $N_r=3$ \\
\hline
25 & 0.0198 & 0.0101 & 0.0268 && 0.0056 & 0.0083 & 0.0116 \\
50 & 0.1408 & 0.2615 & 0.3200 && 0.0107 & 0.0173 & 0.0198 \\
75 & 0.4481 & 0.9036 & 0.8037 && 0.0215 & 0.0556 & 0.0646 \\
90 & 1.4493 & 1.7943 & 1.3402 && 0.3281 & 0.2267 & 0.2416 \\
95 & 4.7151 & 2.2916 & 1.7872 && 0.6050 & 0.5730 & 0.4130 \\
\hline
\end{tabular}
\end{table}

\begin{figure}[!t]
    \centering
    \includegraphics[width=0.92 \columnwidth]{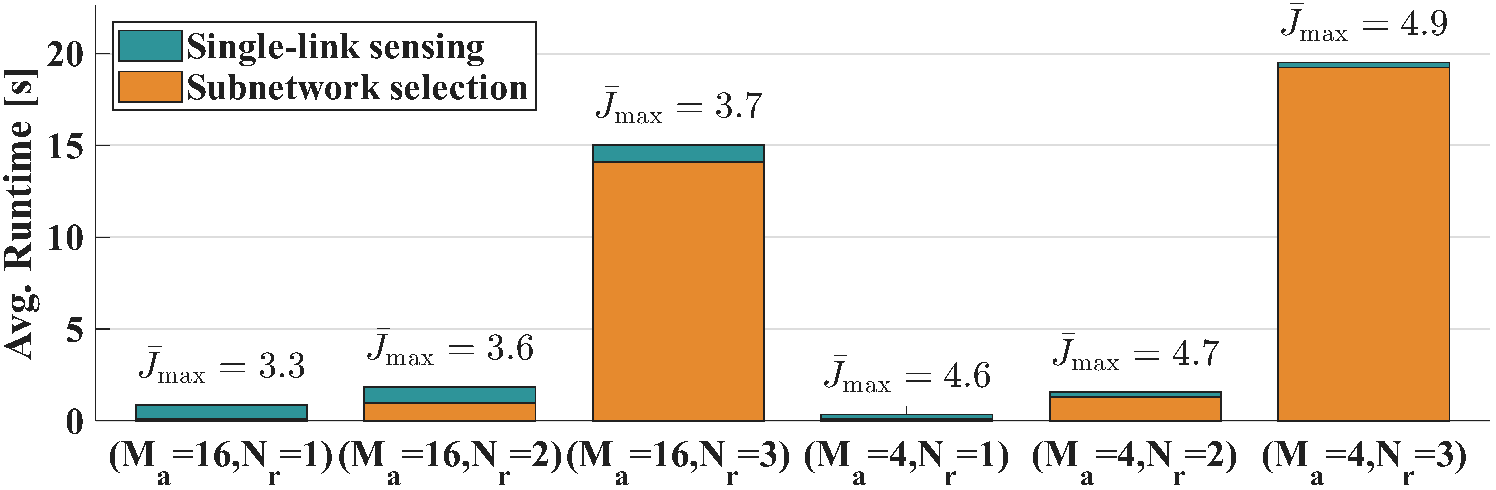}\vspace{-0.3em}
    % \caption{Average runtime of the subnetwork-selection and single-link-sensing stages for different $(M_a, N_r)$ configurations.}
    \caption{Average runtime of the SN-selection and single-link-sensing stages.}
    \label{fig:runtime}
\end{figure}

Fig.~\ref{fig:runtime} reports the simulation runtime until convergence of the single-link-sensing and SN selection stages for different \((M_a,N_r)\) configurations. All results were obtained using MATLAB 2025 on a Windows 11 desktop with an Intel Core Ultra 7 155U CPU at 1.70~GHz, and thus are intended only for relative comparison across configurations, rather than as absolute execution times on optimized software or dedicated hardware. In a real implementation, the sensing-link computations are performed locally at the selected SNs, whereas SN selection and data fusion are performed at the ES. The data-fusion time is omitted because it is negligible compared to the other two stages.
As shown in Fig.~\ref{fig:runtime}, increasing \(M_a\) from 4 to 16 reduces the iterations needed for convergence, but increases the %per-iteration 
single-link sensing time. Hence, for \(N_r=1,2\), the total runtime remains higher for \(M_a=16\), whereas for \(N_r=3\), where SN selection time dominates, the faster convergence of the larger array reduces the total runtime below that of \(M_a=4\). Moreover, increasing \(N_r\) has little effect on single-link-sensing time due to parallel sensing across selected receivers, but substantially increases the ES-side SN-selection time. Together with Table~\ref{tab:percentile_final_error}, this reveals an accuracy-latency trade-off: larger arrays improve localization accuracy, while more receivers mainly improve upper-tail accuracy at the cost of higher latency and resource usage.
With the assumption that a sensing interval spans one OFDM symbol, if all sensing iterations are completed within a single frame, the throughput loss is
\(\eta_{\mathrm{loss}}(\%)=\frac{J_{max}}{140}\times 100 = 0.7143J_{max}\).
%Based on the average number of convergance itertions, the resulting loss is \(2.35\%\), \(2.50\%\), and \(2.64\%\) for \((M_a,N_r)=(16,1)\), \((16,2)\), and \((16,3)\), respectively, and \(3.14\%\), \(3.28\%\), and \(3.28\%\) for \((M_a,N_r)=(4,1)\), \((4,2)\), and \((4,3)\). 
Based on the average number of convergence iterations $\bar{J}_{\max}$, the throughput loss increases from \(2.35\%\) to \(2.64\%\) for \(M_a=16\) and from \(3.28\%\) to \(3.5\%\) for \(M_a=4\) as \(N_r\) increases from 1 to 3.
Together with Table~I, this shows that larger arrays improve localization accuracy while reducing the communication penalty by accelerating convergence. %In contrast, moving from the bistatic setup to the multistatic setup can improve upper-tail localization accuracy, but at the cost of a slight increase in 
In contrast, moving from the bistatic to the multistatic setup improves upper-tail localization accuracy at the expense of a slight increase in
\(\eta_{\mathrm{loss}}\) due to the larger average number of convergence iterations. Nevertheless, the throughput loss remains below \(3.5\%\) for all considered configurations.
%, supporting the orthogonal sensing and communication resource allocation adopted in this paper, %unless the application requires uninterrupted communication service or stringent QoS guarantees.unless uninterrupted communication or strict QoS is required

% \begin{table}[!t]
% \centering
% \caption{sensing-communication-delay}
% \label{tab:cpuA_iter_budget}
% \begin{tabular}{c c c c c c c c}
% \hline
% $(N_{\mathrm{rx}},M)$ & $T_{\mathrm{tot}}$ [ms]   & $J_{\max}(10)$ & Penalty  & $J_{\max}(20)$ & Penalty  \\
% \hline
% $(1,4)$   & 0.621  & 12 & $8.5\%$   & 12 & $4.25\%$ \\
% $(2,4)$     & 0.654& 10 & $ 7.1\%$ & 10 & $3.5\%$ \\
% $(3,4)$    & 1.250 & 7  & $5.00\%$   & 12 & $4.2\%$ \\
% $(1,16)$   & 1.056& 9  & $6.4\%$   & 16 & $ 5.7\%$ \\
% $(2,16)$   & 1.089 & 5  & $3.5\%$   & 5 & $ 1.7\%$ \\
% $(3,16)$  & 1.685  & 5  & $3.5\%$   & 8 & $ 2.8\%$ \\
% \hline
% \end{tabular}
% \end{table}
% To ensure a fair comparison of computational latency and the sensing--communication resource trade-off across the configurations in Fig.~2, all signaling and coordination overhead is absorbed into a common fixed delay \(T_s\). Without loss of generality, we assume \(T_s \approx 0.5\) ms for all configurations. Under this assumption, the variation in total latency is dominated by the configuration-dependent computation time \(T_L\), which is determined mainly by the antenna count and the subset cardinality.  The ES is modeled as a 64-core AMD Ryzen Threadripper PRO 7985WX workstation, while each receiver node is modeled as an 8-core AMD Ryzen 7 8845HS.

\section{Conclusions and Future Work}  \label{Conclusion}
We proposed a novel \ac{ISAC} framework for collaborative multistatic target localization in cluttered \ac{IIoT} environments. The proposed solution periodically learns the background clutter response and suppresses it before target sensing. For target localization, it first uses a sampling-based \ac{FoV}-aware initialization scheme to improve geometric visibility toward the target. Then, an empirical \ac{PEB}-based node refinement scheme is applied to adaptively select the most informative sensing nodes, thus improving localization accuracy with limited resource-latency budget.
%For target localization, it uses a sampling-based \ac{FoV}-aware initialization scheme and an empirical \ac{PEB}-based node refinement scheme to improve geometric visibility toward the target and adaptively select the most informative sensing nodes, thereby improving localization accuracy with limited sensing-resource and latency budgets.
%The \ac{ToA}/\ac{AoA} measurements from the selected links are jointly fused through a reliability-aware \ac{WLS} estimator, where \ac{eCRLB} models weight the reliability of each measurement contribution.
Numerical results show that clutter suppression is essential, with the proposed method achieving near-clutter-free performance. The proposed method reduces the localization RMSE by nearly two orders of magnitude to about \(45\,\mathrm{cm}\), outperforming all considered benchmarks and approaching the oracle best-subset benchmark. Increasing the number of antennas improves localization accuracy and accelerates convergence, while the impact of the number of multistatic receivers is non-monotonic; the communication throughput loss remains below \(3.5\%\) across all considered configurations. Future work will consider adaptive selection of the number of multistatic receivers and joint sensing--communication design under stringent QoS requirements.

\section*{Acknowledgments}
This work is supported by Independent Research Fund Denmark, grant no. 3105-00077B.
\bibliographystyle{IEEEtran}  
\bibliography{refs}

\end{document}